\documentstyle[12pt,epsfig]{article}
\input{epsf}
\newcommand{\be}{\begin{eqnarray}}
\newcommand{\ee}{\end{eqnarray}}
\newcommand{\bea}{\begin{eqnarray}}
\newcommand{\eea}{\end{eqnarray}}
\newcommand{\beq}{\begin{equation}}
\newcommand{\eeq}{\end{equation}}

\newcommand{\lam}{\lambda}

\def\fun#1#2{\lower3.6pt\vbox{\baselineskip0pt\lineskip.9pt
\ialign{$\mathsurround=0pt#1\hfil##\hfil$\crcr#2\crcr\sim\crcr}}}

\begin{document}

\title{Quark-antiquark composite systems:\\
the Bethe--Salpeter equation in the spectral-integration technique
in  case of  different quark masses}
\author{A.V. Anisovich, V.V. Anisovich, V.N. Markov,\\
M.A. Matveev and A.V. Sarantsev }
\date{\today}
\maketitle

\begin{abstract}
The Bethe--Salpeter equations for the quark-antiquark composite
systems with different quark masses, such as $q\bar s$ (with
$q=u$,$d$), $q\bar Q$ and $s \bar Q$ (with $Q=c$,$b$), are written
in terms of  spectral integrals. For  mesons characterized by the
mass $M$, spin $J$ and radial quantum number $n$, the equations
are written for the $(n,M^2)$-trajectories with fixed $J$. The mixing
between states with different quark spin $S$ and angular momentum
$L$ are also discussed.
\end{abstract}

\section{Introduction}

The relativistic description of quark--antiquark states is a necessary
step for meson systematics and the search for exotic states.
The standard way to take account of relativistic effects is to use
the Bethe--Salpeter equation \cite{bethe}. Different versions of the
Bethe--Salpeter equations applied to the description of
quark--antiquark systems may be found in
\cite{petry,petry2,hir,lah,alkof}.
In the present paper
we develop the approach suggested in
\cite{MMM} for the Bethe--Salpeter
equation written for quark--antiquark systems in terms of spectral
integrals. In \cite{MMM},
the systems of quarks with equal masses have been
considered such as $u\bar u, u\bar d, d\bar d, s\bar s$. In this paper,
the systems with unequal masses
 like $q\bar s$
($q=u,d$) and $q\bar Q$ with $Q=c,b$ are treated.

Detailed presentation of the spectral integration method as well as
the emphasis on its advantages was given in \cite{MMM}, so we need not to
repeat ourselves. Let us only stress a particular feature of the
method: it is rather easy to control the quark--gluonium content of the
composite system that is rather important for the search for exotics.
Another advantage consists in the easy treatment of the systems with
high spins.

Similarly to \cite{MMM},
we present here the equations for the group of states
laying on the $(n,M^2)$ trajectory. Such trajectories are linear,
they are suitable for the reconstruction of interaction
between quarks at large distances. We hope, by investigating high-spin
quark--antiquark states, to obtain decisive information on the
structure of forces in the region of $r\sim R_{confinement}$.

In this present paper we present  final formulae for the
$(n,M^2)$-trajectories, the details of the calculations may
be found in \cite{MMM}.
 It should be immediately emphasized that
the case of different masses requires more cumbersome calculations. In
particular, the mixing of states with ($J=L;S=0$) and ($J=L;S=1$)
should be accounted for (here $L$ and $S$ are the orbital moment and
spin of quarks, respectively).

Note that for systems with the equal quark masses we
have already obtained numerical  results for a set of the
$(n,M^2)$-trajectories such as: $a_0,a_1,a_2,\pi,\rho,b_1$.
So we hope that we have elaborated rather
efficient technique allowing us to find
out realistic wave functions for
the quark--antiquark systems. Certain aspects of numerical
solutions of the Bethe--Salpeter equations in terms of the spectral
integrals are discussed in \cite{Markov}.

The paper is organized as follows. In Section 2
we define the quantities
entering the Bethe--Salpeter equation: for equal
masses they were introduced in \cite{MMM},
now we expand the definition for
unequal masses. In Section 3, the equations for the
$(n,M^2)$-trajectories are considered for two different cases,
for ($J=L;S=0$) and ($J=L;S=1$) states. The
technicalities related to the trace calculations of loop diagrams as
well as trace factor convolutions are considered
in Appendices A and B.

\section{Quark-antiquark composite systems}

In the spectral integral technique, the
Bethe--Salpeter equation for the wave function of
the $q_1 \bar q_2$ system
with the total momentum $J$,
angular momentum $L=|\vec J -\vec S|$
and quark-antiquark spin $S$ can be schematically written as:
\be
\left(s-M^2\right)
\widehat\Psi^{(S,L,J)}_{(n)\,\mu_{1}\cdots\mu_{J}} (k_{\perp})=
\label{bs1}
\ee
$$
=\int\frac{d^3 k'_{\perp}}{(2\pi)^3}\Phi(s')
\widehat V\left(s,s',(k_{\perp}k'_{\perp})\right)(\hat k_1'+m_1)
\widehat \Psi^{(S,L,J)}_{(n)\,\mu_{1}
\cdots\mu_{J}} (k'_{\perp})(-\hat k_2'+m_2),
$$
where the
quarks are mass-on-shell: $k_1^2=k_1'^2=m_1^2$ and $k_2^2=k_2'^2=m_2^2$.
We use the following notations:
\be
k=\frac12\left(k_1-k_2\right), \,\,\,\,
P=k_1+k_2 ,
\qquad k'=\frac12\left(k'_1-k'_2\right), \,\,\,\,
P'=k'_1+k'_2,
\label{bs2}
\ee
$$
P^2=s,\ \ P'^2=s', \ \ k^{\perp}_{\mu}=k_{\nu}g^{\perp}_{\nu\mu},
\ \ k'^{\perp}_{\mu}=k'_{\nu}g^{\perp}_{\nu\mu}\, .
$$
The phase space integral is written as
\be
ds'\frac{d^3k'_1}{2k'_{10}}\frac{d^3k'_2}{2k'_{20}}
\delta^{(4)}(P'-k'_1-k'_2)=
\frac{d^3 k'_{\perp}}{(2\pi)^3}\Phi(s'),
\label{bs2a}
\ee
$$
\Phi(s')=\frac{2s' \sqrt{s'}}{s'^2-{(m_1^2-m_2^2)^2}}.
$$
The wave function reads:
\be
\widehat
\Psi^{(S,L,J)}_{(n)\,\mu_{1}\cdots\mu_{J}}(k_{\perp})=
Q^{(S,L,J)}_{\mu_{1}\cdots\mu_{J}}(k_{\perp})\,
\psi^{(S,L,J)}_n (k_{\perp}^2)\ .
\label{bs3}
\ee
Here $Q^{(S,L,J)}_{\mu_{1}\cdots\mu_{J}}(k_{\perp})$
are the moment-operators for fermion-antifermion
systems \cite{operators} defined as follws:
\be
\label{bs10}
&& Q^{(0,J,J)}_{\mu_1\mu_2\ldots\mu_J}(k_{\perp})=i\gamma_5
X^{(J)}_{\mu_1\ldots\mu_J}(k_{\perp}),
\\
&& Q^{(1,J+1,J)}_{\mu_1\ldots\mu_J}(k_{\perp})=\gamma^{\perp}_\alpha
X^{(J+1)}_{\mu_1\ldots\mu_{J}\alpha}(k_{\perp}) ,
\nonumber
\\
&& Q^{(1,J,J)}_{\mu_1\ldots\mu_J}(k_{\perp})=
\varepsilon_{\alpha\nu_1\nu_2\nu_3}\gamma^{\perp}_{\alpha}P_{\nu_1}
Z^{(J)}_{\nu_2\mu_1\ldots\mu_J,\nu_3} (k_{\perp}),
\nonumber
\\
&& Q^{(1,J-1,J)}_{\mu_1\ldots\mu_J}(k_{\perp})=
\gamma^{\perp}_\alpha
Z^{(J-1)}_{\mu_1\ldots\mu_J,\alpha}(k_{\perp}),
\nonumber
\ee
where
\be
&&X^{(J)}_{\mu_1\ldots\mu_J}(k_\perp)
=\frac{(2J-1)!!}{J!}
\bigg [
k^\perp_{\mu_1}k^\perp_{\mu_2}k^\perp_{\mu_3}k^\perp_{\mu_4}
\ldots k^\perp_{\mu_J} -
\label{x-direct}
\\
&&-\frac{k^2_\perp}{2J-1}\left(
g^\perp_{\mu_1\mu_2}k^\perp_{\mu_3}k^\perp_{\mu_4}\ldots k^\perp_{\mu_J}
+g^\perp_{\mu_1\mu_3}k^\perp_{\mu_2}k^\perp_{\mu_4}\ldots
k^\perp_{\mu_J} + \ldots \right)+
\nonumber
\\
&&
+\frac{k^4_\perp}{(2J-1)
(2J-3)}\left(
g^\perp_{\mu_1\mu_2}g^\perp_{\mu_3\mu_4}k^\perp_{\mu_5}
k^\perp_{\mu_6}\ldots k^\perp_{\mu_J}+
\right .
\nonumber
\\
&&
\left .
+g^\perp_{\mu_1\mu_2}g^\perp_{\mu_3\mu_5}k^\perp_{\mu_4}
k^\perp_{\mu_6}\ldots k^\perp_{\mu_J}+
\ldots\right)+\ldots\bigg ]\ ,
\nonumber
\ee
\be
Z^{(J-1)}_{\mu_1\ldots\mu_J, \alpha}(k_\perp)&=&
\frac{2J-1}{L^2}\left (
\sum^J_{i=1}X^{{(J-1)}}_{\mu_1\ldots\mu_{i-1}\mu_{i+1}\ldots\mu_J}
(k_\perp)g^\perp_{\mu_i\alpha} -\right .
\nonumber \\
&&\left . -\frac{2}{2J-1} \sum^J_{i,j=1 \atop i<j}
g^\perp_{\mu_i\mu_j}
X^{{(J-1)}}_{\mu_1\ldots\mu_{i-1}\mu_{i+1}\ldots\mu_{j-1}\mu_{j+1}
\ldots\mu_J\alpha}(k_\perp) \right )\ .
\nonumber
\ee
The potential operator can be represented as a sum of the
$t$-channel operators:
\be
\widehat V\left(s, s', (k_{\perp}k'_{\perp})\right)=
\sum_{I} V^{(0)}_I\left(s, s', (k_{\perp}k'_{\perp})\right)
\widehat O_I \otimes \widehat O_I\ ,
\label{bs4}
\ee
$$
\widehat O_I
={\rm I}, \;\; \gamma_{\mu},\;\; i\sigma_{\mu\nu},\;\;
i\gamma_{\mu}\gamma_5, \;\; \gamma_5 .
$$
To write the spectral integral equations we are to transform
the $t$-channel potential operator
$\widehat V\left(s, s', (k_{\perp}k'_{\perp})\right)$
into the $s$-channel ones as follows:
\be
\widehat V\left(s, s', (k_{\perp}k'_{\perp})\right)=
\sum_I\sum_c  \widehat V^{(0)}_I\left(s, s', (k_{\perp}k'_{\perp})\right)
C_{Ic}\, (\widehat F_c\otimes \widehat F_c),
\label{bs6}
\ee
where $C_{Ic}$ are coefficients of the Fierz matrix:
\be
C=
\left(
   \begin{array}{ccccc}
\frac14 & \frac14  & \frac18  & \frac14  & \frac14   \\ \\
1       & -\frac12 & 0        & \frac12  & -1        \\ \\
3       & 0        & -\frac12 & 0        &  3        \\ \\
1       & \frac12  & 0        & -\frac12 & -1        \\ \\
\frac14 & -\frac14 & \frac18  & -\frac14 & \frac14   \\ \\
   \end{array}
\right).
\ee
Here
the summation  is assumed in the ${i\sigma_{\mu\nu}} \otimes {i\sigma_{\mu\nu}}$
structure for all indices.
Denoting
\be
 V_c\left(s, s', (k_{\perp}k'_{\perp})\right)=
\sum_I  \widehat V^{(0)}_I\left(s, s', (k_{\perp}k'_{\perp})\right)
C_{Ic}\ ,
\label{bs7}
\ee
we have
\be
\widehat V\left(s, s', (k_{\perp}k'_{\perp})\right)=
\sum_c (\widehat F_c\otimes\widehat  F_c)\,
V_c\left(s, s', (k_{\perp}k'_{\perp})\right)=
\label{bs8}
\ee
$$
=({\rm I}\otimes {\rm I})\,V_S\left(s, s', (k_{\perp}k'_{\perp})\right) +
(\gamma_{\mu}\otimes\gamma_{\mu})\,
V_V\left(s, s', (k_{\perp}k'_{\perp})\right) +
(i\sigma_{\mu\nu}\otimes i\sigma_{\mu\nu})\,
\times
\nonumber
$$
$$
\times V_T\left(s, s', (k_{\perp}k'_{\perp})\right)
+(i\gamma_{\mu}\gamma_{\nu}\otimes i\gamma_{\mu}\gamma_{\nu})\,
V_A\left(s, s', (k_{\perp}k'_{\perp})\right)
+(\gamma_5\otimes\gamma_5)\,
V_P\left(s, s', (k_{\perp}k'_{\perp})\right) .
\nonumber
$$
Let us multiply Eq. (\ref{bs1}) by the operator
$ Q^{(S,L,J)}_{\mu_1\ldots\mu_J}(k_{\perp})$
and convolute over the spin-momentum indices:
\be
&&\left(s-M^2\right)
Sp\left[
\widehat \Psi^{(S,L,J)}_{(n)\,\mu_1\ldots\mu_J}(k_{\perp})
(\widehat{k}_1+m_1)
Q^{(S,L,J)}_{\mu_1\ldots\mu_J}(k_{\perp})
(-\widehat{k}_2+m_2)\right]
\\
\nonumber
&&=\sum_{c}
Sp\left[\widehat{F}_c\, (\widehat{k}_1+m_1)
Q^{(S,L,J)}_{\mu_1\ldots\mu_J}(k_{\perp})(-\widehat{k}_2+m_2)\right]
\int\frac{d^3 k_{\perp}'}{(2\pi)^3} \ \Phi(s')
\times
\\
\nonumber
&&\times
V_c\left(s, s', (k_{\perp}k'_{\perp})\right)
Sp\left[(\widehat{k}'_1+m_1)
\widehat{F}_c\, (-\widehat{k}'_2+m_2)
\widehat \Psi^{(S,L,J)}_{(n)\,\mu_1\ldots\mu_J}(k_{\perp}') \right].
\label{bs9}
\ee
We have four states with the $q_1\bar q_2$ spins $S=0$ and $S=1$: \\
1) $S=0$; $L=J$, \\
2) $S=1$; $L=J+1,\,J,\,J-1$, \\
which are mixed and form two final states.
The wave functions read:

for $S=0,1$, $J=L$,
\be
\widehat \Psi^{(S_i,J,J)}_{(n)\,\mu_1\ldots\mu_J}(k_{\perp})=
C_i
\widehat \Psi^{(0,J,J)}_{(n)\,\mu_1\ldots\mu_J}(k_{\perp})
+D_i
\widehat \Psi^{(1,J,J)}_{(n)\,\mu_1\ldots\mu_J}(k_{\perp})
\label{bs.diff.2}
\ee
where $C_i$ and $D_i$ are the mixing coefficients with $i=1,2$.

for $S=1,\, L=J\pm 1,J$,
\be
\widehat \Psi^{(1,(J\pm 1)_j,J)}_{(n)\,\mu_1\ldots\mu_J}(k_{\perp})=
A_j
\widehat \Psi^{(1,J-1,J)}_{(n)\,\mu_1\ldots\mu_J}(k_{\perp})
+B_j
\widehat \Psi^{(1,J+1,J)}_{(n)\,\mu_1\ldots\mu_J}(k_{\perp})\ .
\label{bs15}
\ee
where $A_j$ and $B_j$ are the mixing coefficients with $j=1,2$.

These wave functions are normalized:
\be
\int\frac{d^3k_{\perp}}{(2\pi)^3}\  \Phi(s)
(-1)Sp\left[
\widehat \Psi^{(S',L'_{j'},J')}_{(n')\,\mu_1\ldots\mu_J}(k_{\perp})
(\widehat{k}_1+m_1)
\widehat \Psi^{(S,L_j,J)}_{(n)\,\mu_1\ldots\mu_J}(k_{\perp})
(-\widehat{k}_2+m_2)\right]
\ee
$$
=(-1)^{J} \delta_{S'S}
\delta_{L'_{j'}L_{j}}
\delta_{J'J}\delta_{n'n}\, .
\label{bs12}
$$

\section{Equations for ($n,M^2$) trajectories}
In this section we write the trajectories for ($J=L,S=0,1$)
and ($J=L\pm1,S=1$) states.

\subsection{ The equation for the  ($S=0,1$, $J=L$)-state}

There are  two equations for the two states with $S=0,1$ and
$J=L$. Their wave functions are denoted as
$ C_i \widehat \Psi^{(0,J,J)}_{(n)\,\mu_1\ldots\mu_J}(k_{\perp})
+D_i \widehat \Psi^{(1,J,J)}_{(n)\,\mu_1\ldots\mu_J}(k_{\perp}) $,
with $i=1,2$. These wave functions are orthogonal to each other. Normalization and
orthoganality conditions give three constraints for four
mixing parameters $C_i$ and $D_i$.

Each wave function obeys two equations:
\be
\left(s-M^2\right) \,X^{(J)}_{\mu_1\ldots\mu_J}(k_{\perp})\,
Sp\left[ i\gamma_5\,\gamma^{\perp}_\alpha(\widehat{k}_1+m_1)
i\gamma_5(-\widehat{k}_2+m_2)\right] \times
\label{bs.diff.4}
\ee
$$
\nonumber
\times
\left(
C_i  X^{(J)}_{\mu_1\ldots\mu_{J}}(k_{\perp})\,
        \psi^{(0,J,J)}_n(k_{\perp}^2)+
D_i  \varepsilon_{\alpha \nu_1\nu_2\nu_3} P_{\nu_1}
     Z^{(J)}_{\nu_2 \mu_1\cdots\mu_J,\nu_3}(k_{\perp})\,
    \psi^{(1,J,J)}_n(k_{\perp}^2)\right)=
$$
$$
=X^{(J)}_{\mu_1\ldots\mu_J}(k_{\perp})\,
\sum_c Sp\left[\widehat{F}_c\, (\widehat{k}_1+m_1)
i\gamma_5(-\widehat{k}_2+m_2)\right]  \times
\nonumber
$$
$$
\times\int\frac{d^3 k_{\perp}'}{(2\pi)^3} \  \Phi(s')
V_c \left(s, s', (k_{\perp}k'_{\perp})\right)
Sp\left[ i\gamma_5\, \gamma^{\perp}_{\alpha'} (\widehat{k}'_1+m_1)
\widehat{F}_c\, (-\widehat{k}'_2+m_2)\right]  \times
\nonumber
$$
$$
\times
\left(
C_i  X^{(J)}_{\mu_1\ldots\mu_{J}}(k_{\perp}')\,
        \psi^{(0,J,J)}_n(k_{\perp}'^2)+
D_i  \varepsilon_{\alpha' \nu_1\nu_2\nu_3} P'_{\nu_1}
     Z^{(J)}_{\nu_2 \mu_1\cdots\mu_J,\nu_3}(k_{\perp}')\,
    \psi^{(1,J,J)}_n(k_{\perp}'^2)\right)
$$
and
\be
\left(s-M^2\right) \,\varepsilon_{\beta \nu_1\nu_2\nu_3} P_{\nu_1}
Z^{(J)}_{\nu_2 \mu_1\cdots\mu_J,\nu_3}(k_{\perp})\,
Sp\left[ i\gamma_5\,\gamma^{\perp}_\alpha (\widehat{k}_1+m_1)
\gamma^{\perp}_\beta (-\widehat{k}_2+m_2)\right]  \times
\label{bs.diff.5}
\ee
$$
\nonumber
\times
\left(
C_i  X^{(J)}_{\mu_1\ldots\mu_{J}}(k_{\perp})\,
        \psi^{(0,J,J)}_n(k_{\perp}^2)+
D_i  \varepsilon_{\alpha \nu_1\nu_2\nu_3} P_{\nu_1}
     Z^{(J)}_{\nu_2 \mu_1\cdots\mu_J,\nu_3}(k_{\perp})\,
    \psi^{(1,J,J)}_n(k_{\perp}^2)\right)=
$$
$$
=\varepsilon_{\beta' \nu_1\nu_2\nu_3} P_{\nu_1}
Z^{(J)}_{\nu_2 \mu_1\cdots\mu_J,\nu_3}(k_{\perp})\,
\sum_c Sp\left[\widehat{F}_c\, (\widehat{k}_1+m_1)
\gamma^{\perp}_{\beta'}(-\widehat{k}_2+m_2)\right]\times
\nonumber
$$
$$
\times\int\frac{d^3 k_{\perp}'}{(2\pi)^3} \ \Phi(s')
V_c \left(s, s', (k_{\perp}k'_{\perp})\right)
Sp\left[ i\gamma_5\,\gamma_{\alpha'}^{\perp}  (\widehat{k}'_1+m_1)
\widehat{F}_c\, (-\widehat{k}'_2+m_2)\right] \times
\nonumber
$$
$$
\times
\left(
C_i  X^{(J)}_{\mu_1\ldots\mu_{J}}(k_{\perp}')\,
        \psi^{(0,J,J)}_n(k_{\perp}'^2)+
D_i  \varepsilon_{\alpha' \nu_1\nu_2\nu_3} P'_{\nu_1}
     Z^{(J)}_{\nu_2 \mu_1\cdots\mu_J,\nu_3}(k_{\perp}')\,
    \psi^{(1,J,J)}_n(k_{\perp}'^2)\right)
$$

Now consider the left-hand side of the equation.
Using the traces written in
Appendix A and   convolution of
operators from Appendix B, we have:
$$
X^{(J)}_{\mu_1\ldots\mu_{J}}(k_{\perp})
Sp\left[i\gamma_5 (\widehat{k}_1+m_1)
i\gamma_5(-\widehat{k}_2+m_2)\right]
X^{(J)}_{\mu_1\ldots\mu_{J}}(k_{\perp})=
-2(s-\Delta^2)\alpha (J)k_{\perp}^{2J},
$$
\be
\label{bs.diff.6}
X^{(J)}_{\mu_1\ldots\mu_{J}}(k_{\perp})
Sp\left[\gamma^{\perp}_\alpha (\widehat{k}_1+m_1)
i\gamma_5(-\widehat{k}_2+m_2)\right]
\varepsilon_{\alpha \nu_1\nu_2\nu_3} P_{\nu_1}
Z^{(J)}_{\nu_2\mu_1\ldots\mu_J,\,\nu_3}(k_{\perp})=0
\ee
Here and below we use the following notations:
$\delta=m_2-m_1, \,  \sigma =m_2+m_1$.
Also the left-hand side of (\ref{bs.diff.5}) contains
two convolutions:
\be
\label{bs.diff.7}
\varepsilon_{\beta \nu_1\nu_2\nu_3} P_{\nu_1}
Z^{(J)}_{\nu_2\mu_1\ldots\mu_{J},\,\nu_3}(k_{\perp})
Sp\left[i\gamma_5 (\widehat{k}_1+m_1)
\gamma^{\perp}_\beta(-\widehat{k}_2+m_2)\right]
X^{(J)}_{\mu_1\ldots\mu_{J}}(k_{\perp})=0,
\ee
$$
\varepsilon_{\beta \nu_1\nu_2\nu_3} P_{\nu_1}
Z^{(J)}_{\nu_2\mu_1\ldots\mu_{J},\,\nu_3}(k_{\perp})
Sp\left[\gamma^{\perp}_\alpha (\widehat{k}_1+m_1)
\gamma^{\perp}_\beta(-\widehat{k}_2+m_2)\right]
\times
$$
$$
\times
\varepsilon_{\alpha \nu_1\nu_2\nu_3} P_{\nu_1}
Z^{(J)}_{\nu_2\mu_1\ldots\mu_J,\,\nu_3}(k_{\perp})=-2s(s-\Delta^2)\frac{J(2J+3)^2}{(J+1)^3}\alpha (J)k_{\perp}^{2J}.
$$
The right-hand side of the equation is calculated in two steps.
First, we summarize over $c$:

\be
\label{bs.diff.a}
A\left(s, s', (k_{\perp}k'_{\perp})\right)=
\sum_{c=T,A,P}  A_c\left(s, s', (k_{\perp}k'_{\perp})\right)
V_c\left(s, s', (k_{\perp}k'_{\perp})\right)=
\ee
$$
=
\sum_{c=T,A,P}  Sp\left[\widehat{F}_c\, (\widehat{k}_1+m_1)
i\gamma_5(-\widehat{k}_2+m_2)\right]
Sp\left[i\gamma_5(\widehat{k}'_1+m_1)
\widehat{F}_c\, (-\widehat{k}'_2+m_2)\right]
\times
\nonumber
$$
$$
\times
V_c\left(s, s', (k_{\perp}k'_{\perp})\right)\; ,
$$
\be
\label{bs.diff.b}
B_{\beta'\alpha'}\left (s,s',(k_{\perp}k'_{\perp})\right )=
\sum_{c=T,A,V,S} (B_c)_{\beta'\alpha'}\left(s,s',(k_{\perp}k'_{\perp})\right )
V_c\left (s,s',(k_{\perp}k'_{\perp})\right )=
\ee
$$
=\sum_{c=T,A,V,S} Sp\left[\widehat{F}_c\, (\widehat{k}_1+m_1)
\gamma^{\perp}_{\beta'}(-\widehat{k}_2+m_2)\right]
Sp\left[\gamma^{\perp}_{\alpha'}(\widehat{k}'_1+m_1)
\widehat{F}_c\, (-\widehat{k}'_2+m_2)\right]
\times
$$
$$
\times
 V_c\left (s,s',(k_{\perp}k'_{\perp})\right ),
$$
and
\be
\label{bs.diff.c}
C_{\alpha'}\left(s,s',(k_{\perp}k'_{\perp})\right)=
\sum_{c=T,A} C^c_{\alpha'}\left(s,s',(k_{\perp}k'_{\perp})\right)
V_c\left(s,s',(k_{\perp}k'_{\perp})\right)=
\ee
$$
=\sum_{c=T,A} Sp\left[\widehat{F}_c\, (\widehat{k}_1+m_1)
i\gamma_5(-\widehat{k}_2+m_2)\right]
Sp\left[\gamma^{\perp}_{\alpha'}(\widehat{k}'_1+m_1)
\widehat{F}_c\, (-\widehat{k}'_2+m_2)\right]
\times
$$
$$
\times
V_c\left(s,s',(k_{\perp}k'_{\perp})\right).
$$
In Appendix A the trace calculations are presented, and the values
$A_c\left(s, s', (k_{\perp}k'_{\perp})\right)$, $C_c\left(s, s', (k_{\perp}k'_{\perp})\right)$
are given. So after the summation, $A_c$, $B_c$, $C_c$ are written as follows :
\be
\label{3.1.4}
A\left(s, s', (k_{\perp}k'_{\perp})\right)=
\sum_{c=T,A,P}  A_c\left(s, s', (k_{\perp}k'_{\perp})\right)
V_c\left(s, s', (k_{\perp}k'_{\perp})\right)=
\ee
$$
=-16(k_{\perp}k'_{\perp})\left[2\sqrt{ss'}\,V_T\left(s, s', (k_{\perp}k'_{\perp})\right)
+\Delta^2\, V_A\left(s, s', (k_{\perp}k'_{\perp})\right)\right] -
$$
$$
-4(s-\Delta^2)(s'-\Delta^2)
\left[\frac{\sigma\sigma'}{\sqrt{ss'}}\, V_A\left(s, s', (k_{\perp}k'_{\perp})\right)
+\, V_P\left(s, s', (k_{\perp}k'_{\perp})\right) \right]\, ,
$$
\be
\label{bs.diff.b}
B_{\beta'\alpha'}\left (s,s',(k_{\perp}k'_{\perp})\right )=
\sum_{c=T,A,V,S} (B_c)_{\beta'\alpha'}\left(s,s',(k_{\perp}k'_{\perp})\right )
V_c\left (s,s',(k_{\perp}k'_{\perp})\right ) =
\ee
$$
=4\,g^{\perp}_{\beta'\alpha'}\,
\left[(s-\Delta^2)(s'-\Delta^2)\left(
V_V\left(s,s',(k_{\perp}k'_{\perp})\right)
+2\frac{\sigma^2}{\sqrt{ss'}}\, V_T\left
(s,s',(k_{\perp}k'_{\perp})\right)\right)+ \right . $$ $$ \left .
+4\sqrt{k_{\perp}^2}\sqrt{k_{\perp}'^2}\, z\, \left(\sqrt{ss'}\,
V_A\left (s,s',(k_{\perp}k_{\perp}')\right)
+2\Delta^2\, V_T\left (s,s',(k_{\perp}k'_{\perp})\right)\right)\right]
+ $$
$$ +16\, k^{\perp}_{\beta'}k'^{\perp}_{\alpha'}
\left[\left(\sigma^2\, V_S\left (s,s',(k_{\perp}k'_{\perp})\right)
+\frac{\sigma^2\Delta^2}{\sqrt{ss'}}\,
V_V\left(s,s',(k_{\perp}k'_{\perp})\right)\right)+
\right.
$$
$$
\left.
+4\sqrt{k_{\perp}^2}\sqrt{k_{\perp}'^2}\, z\, V_V\left(s,s',(k_{\perp}k'_{\perp})\right)\right]-
$$
$$
-16\, k'^{\perp}_{\beta'}k^{\perp}_{\alpha'}
\left[\sqrt{ss'}\, V_A\left(s,s',(k_{\perp}k'_{\perp})\right)
+2\Delta^2\, V_T\left(s,s',(k_{\perp}k'_{\perp})\right)\right] +
$$
$$
+16k^{\perp}_{\beta'}k^{\perp}_{\alpha'}\left(s'-\Delta^2\right)
V_V\left(s,s',(k_{\perp}k'_{\perp})\right)
+16k'^{\perp}_{\beta'}k'^{\perp}_{\alpha'}\left(s-\Delta^2\right)
V_V\left(s,s',(k_{\perp}k'_{\perp})\right) \, ,
$$
and
\be
\label{bs.diff.c}
C_{\alpha'}\left(s,s',(k_{\perp}k'_{\perp})\right)=
\sum_{c=T,A} C^c_{\alpha'}\left(s,s',(k_{\perp}k'_{\perp})\right)
V_c\left(s,s',(k_{\perp}k'_{\perp})\right)=
\ee
$$
=8\left[
2\Delta \varepsilon_{\alpha' k k'P'}\, V_A\left(s,s',(k_{\perp}k'_{\perp})\right)
+\sigma \varepsilon_{\alpha' P k'P'}\,
V_A\left(s,s',(k_{\perp}k'_{\perp})\right)+
\right .
$$
$$
\left .
+4\Delta \varepsilon_{\alpha' P k k'}\, V_T\left(s,s',(k_{\perp}k'_{\perp})\right)
+2\sigma \varepsilon_{\alpha' P k P'}\, V_T\left(s,s',(k_{\perp}k'_{\perp})\right)
\right]
$$
Here we used a short notation: \, $ \varepsilon_{\alpha' k k'P'} \equiv k_{\beta} k'_{\mu} P'_{\nu}
\varepsilon_{\alpha' \beta \mu \nu}$.Second , the convolution of operators is performed by using
equations of Appendix B and recurrent formulae for the Legendre
polynomials:
$$
zP_J (z)=\frac{J+1}{2J+1}P_{J+1}(z)+\frac{J}{2J+1}P_{J-1}(z)\, ,
\nonumber
$$
that allows us to represent the Bethe--Salpeter equation in terms of the
Legendre polynomials
.
As a result we get:
\be
X^{(J)}_{\mu_1\ldots\mu_{J}}(k_{\perp})\,
A\left(s,s',(k_{\perp}k'_{\perp})\right)\,
X^{(J)}_{\mu_1\ldots\mu_{J}}(k_{\perp}')=
-4\alpha (J)\left(\sqrt{k_{\perp}^2}\sqrt{k_{\perp}'^2}\right)^{J}
\times
\ee
$$
\times \left[
4\frac{J+1}{2J+1}\sqrt{k_{\perp}^2}\sqrt{k_{\perp}'^2}
\left(2\sqrt{ss'}\,V_T\left(s, s', (k_{\perp}k'_{\perp})\right)
+\Delta^2\,V_A\left(s, s', (k_{\perp}k'_{\perp})\right)
\right)P_{J+1}(z)+
\right .
\nonumber
$$
$$
\left .
+(s-\Delta^2)(s'-\Delta^2)\left(
\frac{\sigma^2}{\sqrt{ss'}}\,V_A\left(s, s',
(k_{\perp}k'_{\perp})\right) +V_P\left(s, s',
(k_{\perp}k'_{\perp})\right) \right)P_J(z)+ \right . \nonumber $$ $$
\left .
+4\frac{J}{2J+1}\sqrt{k_{\perp}^2}\sqrt{k_{\perp}'^2}
\left(2\sqrt{ss'}\, V_T\left(s, s', (k_{\perp}k'_{\perp})\right)
+\Delta^2\,V_A\left(s, s', (k_{\perp}k'_{\perp})\right)
\right)P_{J-1}(z)
\right],
\nonumber
$$

and
\be
X^{(J)}_{\mu_1\ldots\mu_{J}}(k_{\perp})\,
C_{\alpha'}\left (s,s',(k_{\perp}k'_{\perp})\right )\,
\varepsilon_{\alpha' \nu_1\nu_2\nu_3} P'_{\nu_1}
Z^{(J)}_{\nu_2\mu_1\ldots\mu_{J},\,\nu_3}(k_{\perp}')=
\ee
$$
=16\frac{2J+3}{J+1}\alpha (J) \left(\sqrt{k_{\perp}^2}\sqrt{k_{\perp}'^2}\right)^{J+1}
\times
$$
$$
\times\left(\frac{2}{J+1}P'_J(z)-JP_{J+1}(z)\right)
\left(s'\Delta\,V_A\left(s, s', (k_{\perp}k'_{\perp})\right)+
2\Delta\sqrt{ss'}\,V_T\left(s, s', (k_{\perp}k'_{\perp})\right)
\right).
$$
For the right-hand side of (\ref{bs.diff.5}) we have:
\be
\varepsilon_{\beta' \nu_1\nu_2\nu_3} P_{\nu_1}
Z^{(J)}_{\nu_2\mu_1\ldots\mu_{J},\,\nu_3}(k_{\perp})\,
C_{\beta'}\left (s,s',(k_{\perp}k'_{\perp})\right )\,
X^{(J)}_{\mu_1\ldots\mu_{J}}(k_{\perp}')=
\ee
$$
=16\frac{2J+3}{J+1}\alpha (J) \left(\sqrt{k_{\perp}^2}\sqrt{k_{\perp}'^2}\right)^{J+1}
\times
$$
$$
\times\left(\frac{2}{J+1}P'_J(z)-JP_{J+1}(z)\right)
\left(s\Delta\,V_A\left(s, s', (k_{\perp}k'_{\perp})\right)+
2\Delta\sqrt{ss'}\,V_T\left(s, s', (k_{\perp}k'_{\perp})\right) \right)
$$

and
\be
\varepsilon_{\beta' \nu_1\nu_2\nu_3} P_{\nu_1}
Z^{(J)}_{\nu_2\mu_1\ldots\mu_{J},\,\nu_3}(k_{\perp})\,
B_{\beta'\alpha'}\left (s,s',(k_{\perp}k'_{\perp})\right )\,
\varepsilon_{\alpha' \nu_1\nu_2\nu_3} P'_{\nu_1}
Z^{(J)}_{\nu_2\mu_1\ldots\mu_{J},\,\nu_3}(k_{\perp}')=
\ee
$$
=-4\sqrt{ss'}\frac{J(2J+3)^2}{(J+1)^3}
\alpha (J)\left(\sqrt{k_{\perp}^2}\sqrt{k_{\perp}'^2}\right)^{J}
\times
$$
$$
\times \left[
4\frac{J}{2J+1}\sqrt{k_{\perp}^2}\sqrt{k_{\perp}'^2}
\left(\sqrt{ss'}\,V_A\left(s, s', (k_{\perp}k'_{\perp})\right)
+2\Delta^2\,V_T\left(s, s', (k_{\perp}k'_{\perp})\right)
\right)P_{J+1}(z)
\right .
\nonumber
$$
$$
\left .
+(s-\Delta^2)(s'-\Delta^2)\left(
V_V\left(s, s', (k_{\perp}k'_{\perp})\right)
+2\frac{\sigma^2}{\sqrt{ss'}}\,V_T\left(s, s',
(k_{\perp}k'_{\perp})\right) \right)P_J(z)+ \right . \nonumber $$ $$
\left .
+4\frac{J+1}{2J+1}\sqrt{k_{\perp}^2}\sqrt{k_{\perp}'^2}
\left(\sqrt{ss'}\, V_A\left(s, s', (k_{\perp}k'_{\perp})\right)
+2\Delta^2\,V_T\left(s, s', (k_{\perp}k'_{\perp})\right)
\right)P_{J-1}(z)
\right].
\nonumber
$$

Expanding the interaction block in the Legendre polynomials
series,
\be
\label{3.1.7}
V_c\left(s, s', (k_{\perp}k'_{\perp})\right)=
\sum_{J} V^{(J)}_c\left(s, s' \right) P_J(z)=
\ee
$$
=\sum_{J}\widetilde
 V^{(J)}_c\left(s, s' \right)
\alpha(J)\left(-\sqrt{k_{\perp}^2}\sqrt{k_{\perp}'^2}\right)^J P_J(z)\ ,
$$
and integrating over angle variables in the right-hand side
by taking  account of
the standard normalization condition
$\int ^{1}_{-1}dz/2 \;  P^2_J(z)=1/(2J+1)$,
we have finally:
\be
(s-M^2)\left[(s-\Delta^2)\psi^{(0,J,J)}_n(s)C_i \right] =
\ee
$$
=\int\limits_{(m_1+m_2)^2}^{\infty}\frac{ds'}{\pi}\rho(s')
\, 2\, (-k_{\perp}'^2)^{J} \psi^{(0,J,J)}_n(s')C_j
\times
$$
$$
\times \left[
-4\,\xi(J+1)\,\frac{J+1}{2J+1}k_{\perp}^2k_{\perp}'^2
\left(2\sqrt{ss'}\,\widetilde V^{(J+1)}_T\left(s, s'\right)
+\Delta^2\,\widetilde V_A^{(J+1)}\left(s, s'\right)
\right)
+
\right .
$$
$$
\left .
+\xi (J)\,(s-\Delta^2)(s'-\Delta^2)\left(
\frac{\sigma^2}{\sqrt{ss'}}\,
\widetilde V^{(J)}_A\left(s, s'\right)
+\widetilde V^{(J)}_P\left(s, s'\right)
\right)
-
\right .
$$
$$
\left .
-4\,\xi(J-1)\,\frac{J}{2J+1}
\left(2\sqrt{ss'}\, \widetilde V^{(J-1)}_T\left(s, s'\right)
+\Delta^2\,\widetilde V^{(J-1)}_A\left(s, s'\right)
\right)
\right]-
$$
$$
-\int\limits_{(m_1+m_2)^2}^{\infty}\frac{ds'}{\pi}\rho(s')
\, 8\,\frac{2J+3}{J+1}\, k_{\perp}^2\,
(-k_{\perp}'^2)^{J+1}\psi^{(1,J,J)}_n(s')D_j
\times
$$
$$
\times
\left[
\frac{2}{J+1}
\sum_a (2J-4a-1)\,\xi(J-2a-1)\,(-\sqrt{k_{\perp}^2}\sqrt{k_{\perp}'^2})^{-2(a+1)}
\times
\right .
$$
$$
\times\left(s'\Delta\,
\widetilde V^{(J-2a-1)}_A\left(s, s',)\right)
+2\Delta\sqrt{ss'}\,
\widetilde V^{(J-2a-1)}_T\left(s, s',)\right)
\right)
-
$$
$$
-J\,\xi(J+1)
\left(s'\Delta\,\widetilde V^{(J+1)}_A\left(s, s'\right)
+2\Delta'\sqrt{ss'}\,\widetilde V^{(J+1)}_T\left(s, s')\right)
\right),
$$
where
\be
P'_J(z)=\sum_a (2J-4a-1)P_{J-2a-1}(z).
\ee

The second equation, (\ref{bs.diff.5}), reads:
\be
(s-M^2)\left[s(s-\Delta^2)\frac{J(2J+3)^2}{(J+1)^3}
\psi^{(1,J,J)}_n(s)D_i \right] =
\ee
$$
=-\int\limits_{(m_1+m_2)^2}^{\infty}\frac{ds'}{\pi}\rho(s')
\, 8\,\frac{2J+3}{J+1}\, k_{\perp}^2\, (-k_{\perp}'^2)^{J+1}\psi^{(0,J,J)}_n(s')D_j
\times
$$
$$
\times
\left(
\frac{2}{J+1}
\sum_a (2J-4a-1)\,\xi(J-2a-1)\,(-\sqrt{k_{\perp}^2}\sqrt{k_{\perp}'^2})^{-2(k+1)}
\times
\right .
$$
$$
\times\left(s\Delta\,
\widetilde V^{(J-2a-1)}_A\left(s, s',)\right)
+2\Delta\sqrt{ss'}\,
\widetilde V^{(J-2a-1)}_T\left(s, s')\right)
\right)-
$$
$$
-J\,\xi(J+1)
\left(s\Delta\,\widetilde V^{J+1}_A\left(s, s'\right)
+2\Delta\sqrt{ss'}\,\widetilde V^{J+1}_T\left(s, s')\right)
\right)+
$$
$$
+\int\limits_{(m_1+m_2)^2}^{\infty}\frac{ds'}{\pi}\rho(s')
\, 2\, \sqrt{ss'} \frac{J(2J+3)^2}{(J+1)^3}(-k_{\perp}'^2)^{J}
\psi^{(1,J,J)}_n(s')D_j\times
$$
$$
\times \left[
-4\,\xi(J+1)\,\frac{J}{2J+1}k_{\perp}^2k_{\perp}'^2
\left(\sqrt{ss'}\,\widetilde V^{(J+1)}_A\left(s, s'\right)
+2\Delta^2\,\widetilde V^{(J+1)}_T\left(s, s'\right)
\right)
+
\right.
$$
$$
\left .
+\xi (J)\,(s-\Delta^2)(s'-\Delta^2)\left(
2\frac{\sigma^2}{\sqrt{ss'}}\,
\widetilde V^{(J)}_T\left(s, s'\right)
+\widetilde V^{(J)}_V\left(s, s'\right)
\right)
-
\right .
$$
$$
\left .
-4\,\xi(J-1)\,\frac{J+1}{2J+1}
\left(\sqrt{ss'}\, \widetilde V^{(J-1)}_A\left(s, s'\right)
+2\Delta^2\,\widetilde V^{(J-1)}_T\left(s, s'\right)
\right)
\right].
$$
The normalization and orthogonality
conditions look as follows:
\be
\int\limits_{(m_2+m_1)^2}^{\infty}\frac{ds}{\pi}\rho(s)
\left [
C^2_i \left( \psi^{(0,J,J)}_n(k_{\perp}^2)\right)^2
2\alpha (J)(-k_{\perp}^2)^{J}\left(s-\Delta^2\right) +
\right .
\ee
$$
\left .
+
D_i^2 \left( \psi^{(1,J,J)}_n(k_{\perp}^2) \right)^2
2\alpha(J)(-k_{\perp}^2)^{J}s\left(s-\Delta^2\right)\frac{J(2J+3)^2}{(J+1)^3}
\right] =1, \qquad i=1,2,
$$
and
\be
\int\limits_{(m_2+m_1)^2}^{\infty}\frac{ds}{\pi}\rho(s)
\left [
C_1C_2 \left( \psi^{(0,J,J)}_n(k_{\perp}^2)\right)^2
2\alpha (J)(-k_{\perp}^2)^{J}\left(s-\Delta^2\right)   +
\right .
\ee
$$
\left .
+
D_1D_2 \left( \psi^{(1,J,J)}_n(k_{\perp}^2) \right)^2
2\alpha(J)(-k_{\perp}^2)^{J}s\left(s-\Delta^2\right)\frac{J(2J+3)^2}{(J+1)^3}
\right] =0.
$$

\subsection{The equations for the $(S=1,J=L\pm 1)$-states}

We have two equations for the two states with $S=1$ and
$J=L\pm 1$. Their wave functions are denoted as
$ A_j \widehat \Psi^{(1,J-1,J)}_{(n)\,\mu_1\ldots\mu_J}(k_{\perp})
+B_j \widehat \Psi^{(1,J+1,J)}_{(n)\,\mu_1\ldots\mu_J}(k_{\perp})
$, with $j=1,2$. These wave functions are orthogonal. Normalization and
orthogonality conditions give three constraints for four
mixing parameters $A_j$ and $B_j$.

Each wave function obeys two equations:
\be
\label{bs.diff.8}
\left(s-M^2\right)
X^{(J+1)}_{\mu_1\ldots\mu_{J}\beta}(k_{\perp})
Sp\left[\gamma^{\perp}_\alpha (\widehat{k}_1+m_1)
\gamma^{\perp}_\beta(-\widehat{k}_2+m_2)\right]    \times
\ee
$$
\nonumber
\times
\left(
A_j  Z^{(J-1)}_{\mu_1\ldots\mu_J,\alpha}(k_{\perp})\,
        \psi^{(1,J-1,J)}_n(k_{\perp}^2) +
B_j  X^{(J+1)}_{\mu_1\ldots\mu_{J}\alpha}(k_{\perp})\,
        \psi^{(1,J+1,J)}_n(k_{\perp}^2) \right)=
$$
$$
\nonumber
=X^{(J+1)}_{\mu_1\ldots\mu_J\beta'}(k_{\perp})
\sum_c Sp\left[\widehat{F}_c (\widehat{k}_1+m_1)
\gamma^{\perp}_{\beta'} (-\widehat{k}_2+m_2)\right]  \times
$$
$$
\nonumber
\times
\int\frac{d^3 k_{\perp}'}{(2\pi)^3 }
V_c \left(s,s', (k_{\perp}k'_{\perp})\right)
Sp\left[\gamma^{\perp}_{\alpha'}(\widehat{k}'_1+m_1)
\widehat{F}_c(-\widehat{k}'_2+m_2)\right]
\times
$$
$$
\nonumber
\times
\left(
A_j  Z^{(J-1)}_{\mu_1\ldots\mu_J,\,\alpha'}(k_{\perp}')\,
        \psi^{(1,J-1,J)}_n(k_{\perp}'^2)  +
B_j  X^{(J+1)}_{\mu_1\ldots\mu_J\alpha'}(k_{\perp}')\,
        \psi^{(1,J+1,J)}_n(k_{\perp}'^2)\right),
$$
and
\be
\label{bs.diff.9}
\left(s-M^2\right)
Z^{(J-1)}_{\mu_1\ldots\mu_J,\beta}(k_{\perp})\,
Sp\left[\gamma^{\perp}_\alpha (\widehat{k}_1+m_1)
\gamma^{\perp}_\beta(-\widehat{k}_2+m_2)\right] \times
\ee
$$
\nonumber
\times
\left(
A_j  Z^{(J-1)}_{\mu_1\ldots\mu_J,\alpha}(k_{\perp})\,
        \psi^{(1,J-1,J)}_n(k_{\perp}^2) +
B_j X^{(J+1)}_{\mu_1\ldots\mu_{J}\alpha}(k_{\perp})\,
        \psi^{(1,J+1,J)}_n(k_{\perp}^2) \right)=
$$
$$
\nonumber
=Z^{(J-1)}_{\mu_1\ldots\mu_J,\beta'}(k_{\perp})\,
\sum_c Sp\left[\widehat{F}_c (\widehat{k}_1+m_1)
\gamma^{\perp}_{\beta'} (-\widehat{k}_2+m_2)\right]  \times
$$
$$
\nonumber
\times
\int\frac{d^3 k_{\perp}'}{(2\pi)^3 }
V_c \left( s,s', (k_{\perp}k'_{\perp})\right)
Sp\left[\gamma^{\perp}_{\alpha'}(\widehat{k}'_1+m_1)
\widehat{F}_c(-\widehat{k}'_2+m_2)\right]   \times
$$
$$
\nonumber
\times
\left(
A_j  Z^{(J-1)}_{\mu_1\ldots\mu_J,\,\alpha'}(k_{\perp}')\,
        \psi^{(1,J-1,J)}_n(k_{\perp}'^2)  +
B_j  X^{(J+1)}_{\mu_1\ldots\mu_J\alpha'}(k_{\perp}')\,
        \psi^{(1,J+1,J)}_n(k_{\perp}'^2)\right).
$$
Firstly, consider (\ref{bs.diff.8}).
In the left-hand side of (\ref{bs.diff.8})
one has two convolutions:
\be
\label{bs.diff.10}
X^{(J+1)}_{\mu_1\ldots\mu_{J}\beta}(k_{\perp})
Sp\left[\gamma^{\perp}_\alpha (\widehat{k}_1+m_1)
\gamma^{\perp}_\beta(-\widehat{k}_2+m_2)\right]
X^{(J+1)}_{\mu_1\ldots\mu_{J}\alpha}(k_{\perp}) =
\ee
$$
=2\alpha (J)k_{\perp}^{2(J+1)}\left[\frac{2J+1}{J+1}(s-\Delta^2)+4k_{\perp}^2\right],
$$
$$
X^{(J+1)}_{\mu_1\ldots\mu_{J}\beta}(k_{\perp})
Sp\left[\gamma^{\perp}_\alpha (\widehat{k}_1+m_1)
\gamma^{\perp}_\beta(-\widehat{k}_2+m_2)\right]
Z^{(J-1)}_{\mu_1\ldots\mu_{J},\,\alpha}(k_{\perp})=
8\alpha (J)k_{\perp}^{2(J+1)}.
$$
Also the left-hand side of (\ref{bs.diff.9}) contains
two convolutions:
\be
\label{bs.diff.11}
Z^{(J-1)}_{\mu_1\ldots\mu_{J},\beta}(k_{\perp})
Sp\left[\gamma^{\perp}_\alpha (\widehat{k}_1+m_1)
\gamma^{\perp}_\beta(-\widehat{k}_2+m_2)\right]
X^{(J+1)}_{\mu_1\ldots\mu_{J}\alpha}(k_{\perp})
=8\alpha (J) k_{\perp}^{2(J+1)},
\ee
$$
Z^{(J-1)}_{\mu_1\ldots\mu_{J},\beta}(k_{\perp})
Sp\left[\gamma^{\perp}_\alpha (\widehat{k}_1+m_1)
\gamma^{\perp}_\beta(-\widehat{k}_2+m_2)\right]
Z^{(J-1)}_{\mu_1\ldots\mu_{J},\alpha}(k_{\perp})=
$$
$$
=2\alpha (J)k_{\perp}^{2(J-1)}\left[\frac{2J+1}{J}(s-\Delta^2)+4k_{\perp}^2\right].
$$

The right-hand side of Eqs. (\ref{bs.diff.8})
and (\ref{bs.diff.9}) is determined by the convolutions
of the trace factor
$ B_{\beta'\alpha'}\left (s,s',(k_{\perp}k'_{\perp})\right )$,
see Eqs. (\ref{bs.diff.b}) , with
angular-momentum wave functions; corresponding
formulae may be found  in Appendix B. Taking them into account
one has for the right-hand side of (\ref{bs.diff.8}):
\be
X^{(J+1)}_{\mu_1\ldots\mu_{J}\beta'}(k_{\perp})
\, B_{\beta'\alpha'}\left (s,s',(k_{\perp}k'_{\perp})\right )\,
X^{(J+1)}_{\mu_1\ldots\mu_{J}\alpha'}(k_{\perp}')=
4\alpha (J)\left(\sqrt{k_{\perp}^2}\sqrt{k_{\perp}'^2}\right)^{J+1}
\times
\ee
$$
\times\left(\left[
\frac{2J+1}{J+1}(s-\Delta^2)(s'-\Delta^2)
\left(V_V (s,s',(k_{\perp}k'_{\perp}))
+2\frac{\sigma^2}{\sqrt{ss'}}\, V_T
(s,s',(k_{\perp}k'_{\perp}))\right)+
\right .\right .
$$
$$
\left .
+4(s'-\Delta^2)k_{\perp}^2\, V_V (s,s',(k_{\perp}k'_{\perp}))
+4(s-\Delta^2)k_{\perp}'^2\,  V_V (s,s',(k_{\perp}k'_{\perp}) )+
\right.
$$
$$
\left.
+16\frac{J+1}{2J+1}k_{\perp}^2k_{\perp}'^2\, V_V (s,s',(k_{\perp}k_{\perp}'))\right] P_{J+1}(z)
+
$$
$$
+4\left[\sigma^2\, V_S (s,s',(k_{\perp}k'_{\perp}) )
+\frac{\sigma^2\Delta^2}{\sqrt{ss'}}\,
V_V (s,s',(k_{\perp}k'_{\perp}))+
\right.
$$
$$
\left .
+\frac{J}{J+1}\left(\sqrt{ss'}\, V_A (s,s',(k_{\perp}k'_{\perp}))
+2\Delta^2 V_T (s,s',(k_{\perp}k'_{\perp}))\right)
\right]
\sqrt{k_{\perp}^2}\sqrt{k_{\perp}'^2}P_{J}(z)+
$$
$$
\left .
+16\frac{J}{2J+1}k_{\perp}^2k'^2\, V_V (s,s',(k_{\perp}k'_{\perp}))  P_{J-1}(z) \right),
$$

and
\be
X^{(J+1)}_{\mu_1\ldots\mu_{J}\beta'}(k_{\perp})
\, B_{\beta'\alpha'}\left (s,s',(k_{\perp}k'_{\perp})\right )\,
Z^{(J-1)}_{\mu_1\ldots\mu_{J},\,\alpha'}(k_{\perp}')=
16\alpha (J) k_{\perp}^2\left(\sqrt{k_{\perp}^2}\sqrt{k_{\perp}'^2}\right)^{J-1}
\times
\ee
$$
\times\left(
\left[s-\Delta^2+4\frac{J+1}{2J+1}k_{\perp}^2\right]k_{\perp}'^2\,
V_V (s,s',(k_{\perp}k'_{\perp}))\, P_{J+1}(z)+ \right .
$$
$$
+\left[\sigma^2\, V_S(s,s',(k_{\perp}k'_{\perp}))
+\frac{\sigma^2\Delta^2}{\sqrt{ss'}}\,V_V
(s,s',(k_{\perp}k'_{\perp}))-
\right .
$$
$$
\left .
-\sqrt{ss'}\, V_A(s,s',(k_{\perp}k'_{\perp}))
-2\Delta^2\, V_T(s,s',(k_{\perp}k'_{\perp}))
\right]
\sqrt{k_{\perp}^2}\sqrt{k_{\perp}'^2}P_{J}(z)+
$$
$$
\left .
+\left[s'-\Delta^2+4\frac{J}{2J+1}k_{\perp}'^2\right]k_{\perp}^2
\, V_V(s,s',(k_{\perp}k'_{\perp}))\, P_{J-1}(z)
\right) .
$$
And for the right side of (\ref{bs.diff.9}):
\be
Z^{(J-1)}_{\mu_1\ldots\mu_{J},\beta'}(k_{\perp})
\, B_{\beta'\alpha'}\left (s,s',(k_{\perp}k'_{\perp})\right )\,
X^{(J+1)}_{\mu_1\ldots\mu_{J}\alpha'}(k_{\perp}')=
16\alpha (J)k_{\perp}'^2\left(\sqrt{k_{\perp}^2}\sqrt{k_{\perp}'^2}\right)^{J-1}
\times
\ee
$$
\times\left(
\left[s'-\Delta^2+4\frac{J+1}{2J+1}k_{\perp}'^2\right]k_{\perp}^2\,
V_V (s,s',(k_{\perp}k'_{\perp}))\, P_{J+1}(z)+ \right .
$$
$$
+\left[\sigma^2\, V_S(s,s',(k_{\perp}k'_{\perp}))
+\frac{\sigma^2\Delta^2}{\sqrt{ss'}}\,V_V
(s,s',(k_{\perp}k'_{\perp}))-
\right .
$$
$$
\left .
-\sqrt{ss'}\, V_A(s,s',(k_{\perp}k'_{\perp}))
-2\Delta^2\, V_T(s,s',(k_{\perp}k'_{\perp}))
\right]
\sqrt{k_{\perp}^2}\sqrt{k_{\perp}'^2}P_{J}(z)+
$$
$$
\left .
+\left[s-\Delta^2+4\frac{J}{2J+1}k_{\perp}^2\right]k_{\perp}'^2
\, V_V(s,s',(k_{\perp}k'_{\perp}))\, P_{J-1}(z)
\right) .
$$

and
\be
Z^{(J-1)}_{\mu_1\ldots\mu_{J},\beta'}(k_{\perp})
\, B_{\beta'\alpha'}\left (s,s',(k_{\perp}k'_{\perp})\right )\,
Z^{(J-1)}_{\mu_1\ldots\mu_{J},\alpha'}(k_{\perp}')=
4\alpha (J)\left(\sqrt{k_{\perp}^2}\sqrt{k_{\perp}'^2}\right)^{J-1}
\times
\ee
$$
\times\left(
16\frac{J+1}{2J+1}k_{\perp}^2k_{\perp}'^2\, V_V(s,s',(k_{\perp}k_{\perp}'))\,
P_{J+1}(z)+
\right .
$$
$$
+4\left[\,\sigma^2 V_S(s,s',(k_{\perp}k'_{\perp}))
+\frac{\sigma^2\Delta^2}{\sqrt{ss'}}\,V_V
(s,s',(k_{\perp}k'_{\perp}))+
\right .
$$
$$
\left .
+\frac{J+1}{J}\left(\sqrt{ss'}\, V_A (s,s',(k_{\perp}k'_{\perp}))
+2\Delta^2 V_T (s,s',(k_{\perp}k'_{\perp}))\right)
\right]
\sqrt{k_{\perp}^2}\sqrt{k_{\perp}'^2}P_{J}(z)+
$$
$$
+\left(\left[
\frac{2J+1}{J}(s-\Delta^2)(s'-\Delta^2)
\left(V_V (s,s',(k_{\perp}k'_{\perp}))
+2\frac{\sigma^2}{\sqrt{ss'}}\, V_T
(s,s',(k_{\perp}k'_{\perp}))\right)+
\right .\right .
$$
$$
\left .
+4(s'-\Delta^2)k_{\perp}^2\, V_V (s,s',(k_{\perp}k'_{\perp}))
+4(s-\Delta^2)k_{\perp}'^2\,  V_V (s,s',(k_{\perp}k'_{\perp}) )+ \right.
$$
$$
\left.
+16\frac{J}{2J+1}k_{\perp}^2k_{\perp}'^2\, V_V (s,s',(k_{\perp}k_{\perp}'))\right]
P_{J-1}(z).
$$

In the right-hand sides of eqs. (\ref{bs.diff.8}) and (\ref{bs.diff.9}),
we expand the interaction block in the Legendre polinomials
series and integrate over angle variables
$\int ^{1}_{-1} dz/2$.
As a result, Eq. (\ref{bs.diff.8}) reads:
\be
(s-M^2) \left[ 4\psi^{(1,J-1,J)}_n(s)A_j  +
      \left(\frac{2J+1}{J+1}(s-\Delta^2)+4k_{\perp}^2\right)
        \psi^{(1,J+1,J)}_n(s)B_j\right]=
\ee
$$
\nonumber
=\int\limits_{(m_1+m_2)^2}^{\infty}\frac{ds'}{\pi}\rho(s')
\, 8\, (-k_{\perp}'^2)^{J-1}
 \psi^{(1,J-1,J)}_n(s')A_j
 \times
$$
$$
\nonumber
\times
\left[
     \xi (J+1)\,k_{\perp}'^4
           \left(s-\Delta^2+4\frac{J+1}{2J+1}k_{\perp}^2 \right)
            \, \widetilde V^{(J+1)}_V(s,s')
            -
\right .
$$
$$
-\xi (J)\, k_{\perp}'^2
\left[\sigma^2\,\widetilde  V^{(J)}_S(s,s')
+\frac{\sigma^2\Delta^2}{\sqrt{ss'}}\,
\widetilde V^{(J)}_V (s,s')-
\right .
$$
$$
\left .
-\sqrt{ss'}\, \widetilde  V^{(J)}_A(s,s')
-2\Delta^2\, \widetilde  V^{(J)}_T(s,s')
\right]+
$$
$$
\left .
+\xi (J-1)\,\left[s'-\Delta^2+4\frac{J}{2J+1}k_{\perp}'^2\right]
\, \widetilde  V^{(J-1)}_V(s,s')
\right)+
$$
$$
+\int\limits_{(m_1+m_2)^2}^{\infty}\frac{ds'}{\pi}\rho(s')
\, 2\, (-k_{\perp}'^2)^{J+1}
 \psi^{(1,J+1,J)}_n(k_{\perp}'^2)B_j
 \times
$$
$$
\times
\left( \xi(J+1)\,
\left[\frac{2J+1}{J+1}(s-\Delta^2)(s'-\Delta^2)
\left( \widetilde  V^{(J+1)}_V(s,s')
+2\frac{\sigma^2}{\sqrt{ss'}}\,
\widetilde  V^{(J+1)}_T(s,s') \right)+
\right .\right .
$$
$$
+4(s'-\Delta^2)k_{\perp}^2\, \widetilde  V^{(J+1)}_V(s,s')
+4(s-\Delta^2)k_{\perp}'^2\,  \widetilde  V^{(J+1)}_V(s,s')+
$$
$$
\left .
+16\frac{J+1}{2J+1}k_{\perp}^2k_{\perp}'^2\,
\widetilde  V^{(J+1)}_V(s,s')\right]-
$$
$$
-4\,\xi (J)\,
\left[\sigma^2\, \widetilde  V^{(J)}_S(s,s')
+\frac{\sigma^2\Delta^2}{\sqrt{ss'}}\,
\widetilde  V^{(J)}_V(s,s')+
\right .
$$
$$
\left .
+\frac{J}{J+1}\left(\sqrt{ss'}\, \widetilde  V^{(J)}_A(s,s')
+2\Delta^2 \widetilde  V^{(J)}_T(s,s')\right)
\right]+
$$
$$
\left .
+16\,\xi (J-1)\,\frac{J}{2J+1}\,
\widetilde  V^{(J-1)}_V(s,s') \right).
$$

The second equation, (\ref{bs.diff.9}), reads:

\be
(s-M^2)  \left[ \left(\frac{2J+1}{J}(s-\Delta^2)+4k_{\perp}^2\right)
                  \psi^{(1,J-1,J)}_n(s)A_j
             + 4k_{\perp}^4 \psi^{(1,J+1,J)}_n(s)B_j\right]=
\ee
$$
\nonumber
=\int\limits_{(m_1+m_2)^2}^{\infty}\frac{ds'}{\pi}\rho(s')
2\, (-k_{\perp}'^2)^{J-1}
 \psi^{(1,J-1,J)}_n(s')A_j
 \times
$$
$$
\times\left(
16\, \xi (J+1)\,\frac{J+1}{2J+1}k_{\perp}^4k_{\perp}'^4\,
\widetilde  V^{(J+1)}_V(s,s')-
\right .
$$
$$
-4\,\xi (J)\,k_{\perp}^2k_{\perp}'^2\left[\,\sigma^2 \widetilde
V^{(J)}_S(s,s') +\frac{\sigma^2\Delta^2}{\sqrt{ss'}}\,
\widetilde  V^{(J)}_V(s,s')+
\right .
$$
$$
\left .
+\frac{J+1}{J}\left(\sqrt{ss'}\, \widetilde  V^{(J)}_A(s,s')
+2\Delta^2 \widetilde  V^{(J)}_T(s,s')\right)
\right]-
$$
$$
+\xi (J-1)\,\left(\left[
\frac{2J+1}{J}(s-\Delta^2)(s'-\Delta^2)
\left(\widetilde  V^{(J-1)}_V(s,s')
+2\frac{\sigma^2}{\sqrt{ss'}}\,
\widetilde  V^{(J-1)}_T(s,s')\right)+
\right .\right .
$$
$$
\left .
+4(s'-\Delta^2)k_{\perp}^2\, \widetilde  V^{(J-1)}_V(s,s')
+4(s-\Delta^2)k_{\perp}'^2\,  \widetilde  V^{(J-1)}_V(s,s')
+
\right.
$$
$$
\left.
+16\frac{J}{2J+1}k_{\perp}^2k_{\perp}'^2\, \widetilde  V^{(J-1)}_V(s,s')
\right]+
$$
$$
\nonumber
+\int\limits_{(m_1+m_2)^2}^{\infty}\frac{ds'}{\pi}\rho(s')
\, 8\, (-k_{\perp}'^2)^{J+1} \psi^{(1,J+1,J)}_n(s')B_j
\times
$$
$$
\times \xi (J+1)\, k_{\perp}^4 \left(
\left[s'-\Delta^2+4\frac{J+1}{2J+1}k_{\perp}'^2\right]\,
\widetilde  V^{(J+1)}_V(s,s')
-
\right .
$$
$$
-\xi (J)\,k_{\perp}^2\left[\sigma^2\,
\widetilde  V^{(J)}_S(s,s')
+\frac{\sigma^2\Delta^2}{\sqrt{ss'}}\,
\widetilde  V^{(J)}_V(s,s')
-
\right .
$$
$$
\left .
-\sqrt{ss'}\, \widetilde  V^{(J)}_A(s,s')
-2\Delta^2\, \widetilde  V^{(J)}_T(s,s')
\right]
+
$$
$$
\left .
+\xi (J-1)\,\left[s-\Delta^2+4\frac{J}{2J+1}k_{\perp}^2\right]
\, \widetilde  V^{(J-1)}_V(s,s')
\right) .
$$

Normalization and orthogonality
conditions are :
\be
\int\limits_{(m_2+m_1)^2}^{\infty}\frac{ds}{\pi}\rho(s)
\left [
A^2_j \left( \psi^{(1,J-1,J)}_n(k_{\perp}^2)\right)^2
2\alpha (J)(-k_{\perp}^2)^{(J-1)}
\times
\right .
\ee
$$
\left .
\times
\left(\frac{2J+1}{J}(s-\Delta^2)+4k_{\perp}^2\right)+
\right.
$$
$$
\left.
+2A_j B_j  \psi^{(1,J-1,J)}_n(k_{\perp}^2)\psi^{(1,J+1,J)}_n(k_{\perp}^2)
8\alpha (J) (-k_{\perp}^2)^{(J+1)}+
\right .
$$
$$
\left .
+
B_j^2 \left( \psi^{(1,J+1,J)}_n(k_{\perp}^2) \right)^2
2\alpha(J)(-k_{\perp}^2)^{(J+1)}\left(\frac{2J+1}{J+1}(s-\Delta^2)+4k_{\perp}^2\right)
\right] =1 \qquad j=1,2,
$$
and
\be
\int\limits_{(m_2-m_1)^2}^{\infty}\frac{ds}{\pi}\rho(s)
\left [
A_1 A_2\left( \psi^{(1,J-1,J)}_n(k_{\perp}^2)\right)^2
2\alpha (J)(-k_{\perp}^2)^{(J-1)}
\times
\right .
\ee
$$
\left .
\times
\left(\frac{2J+1}{J}(s-\Delta^2)+4k_{\perp}^2\right)+
\right .
$$
$$
\left.
+(A_1 B_2+A_2 B_1)   \psi^{(1,J-1,J)}_n(k_{\perp}^2)\psi^{(1,J+1,J)}_n(k_{\perp}^2)
8\alpha (J) (-k_{\perp}^2)^{(J+1)}+
\right .
$$
$$
\left .
+
B_1B_2 \left( \psi^{(1,J+1,J)}_n(k_{\perp}^2) \right)^2
2\alpha (J)(-k_{\perp}^2)^{(J+1)}\left( \frac{2J+1}{J+1}(s-\Delta^2)+4k_{\perp}^2
\right) \right] =0.
$$

Let us emphasize again: all the above equations are written for  $J>0$.

\section{Conclusion}

We have presented the Bethe-Salpeter equations for the quark-antiquark
systems when the quark and antiquark have different masses.
The main difference from the equal mass case is that there is the mixture of states ($J=L$, $S=0$)
and ($J=L$, $S=1$) and  that  is proportional to the quark mass difference.
The mixing between  ($S=1$, $J=L$) and
($S=0$, $J=L$) states give rise to a  strongly correlated
system of equations. In the equation for states with total spin $J$
we need to know all lower projections of the potential on the
Legendre polynomials, not only  the $J+1$, $J$ ,$J-1$ ones.
The numerical study of this equations is now in progress.

We are grateful to  L.G. Dakhno, M.N. Kobrinsky, Y.S. Kalashnikova, B.Ch. Metsch, 
V.A. Nikonov , H.R. Petry and V.V Vereshagin  for stimulating and
useful discussions. The paper is supported by the RFBR grant no
01-02-17861. One of us (V.N.M.) is supported in part by
INTAS call 2000 project 587 , RFBR 1-02-17152 and "Dynasty Foundation".

\section{Appendix A: Traces for loop diagrams}

Here we present the traces  used in the calculation of
loop diagrams. Recall that  in the spectral integral representation, there
is no energy conservation, $s\neq s'$, where $P^2=s$, $P'^2=s'$, but
all constituents are mass-on-shell:
$$
k^2_1=m^2_1,\,\,\,k^2_2=m^2_2,\,\,\,k'^2_1=m^2_1,\,\,\,k'^2_2=m^2_2.
$$
We have used notations for the quark momenta:
\be
&& k_{\nu}=\frac12 (k_1-k_2)_{\nu} ,\,\,\,
k'_{\nu}=\frac12 (k'_1-k'_2)_{\nu}\, ,
\\
\nonumber
&& k^{\perp}_{\mu}=k_{\nu}g^{\perp}_{\nu\mu},\,\,\,
k'^{\perp}_{\mu}=k'_{\nu}g^{\perp}_{\nu\mu}\, ,
\\
\nonumber
&& k_{\mu}=\frac{m^2_1-m^2_2}{2s}P_{\mu}+k^{\perp}_{\mu},\,\,\,
k'_{\mu}=\frac{m^2_1-m^2_2}{2s'}P'_{\mu}+k'^{\perp}_{\mu}\, ,
\ee
and for the quarks masses :
\be
\Delta=m_2-m_1,\,\, \,  \sigma=m_2+m_1,\,.
\ee
We work  with the following definition of the matrices:
$$
\gamma_5=-i\gamma^0 \gamma^1 \gamma^2 \gamma^3,\,\,\,\,
\sigma_{\mu\nu}=\frac12 \left[\gamma_{\mu}\gamma_{\nu}\right].
$$

\subsection{Traces for the $S=0$ states}

For  the $S=0$ states we have the following non-zero traces:
\be
&&T'_P=Tr\left[i\gamma_5
(\hat k'_1+m_1)\gamma_5(-\hat k'_2+m_2)\right]=2i(s'-(m_2-m_1)^2)  ,
\\
\nonumber
&&T'_A=Tr\left[i\gamma_5
(\hat k'_1+m_1)i\gamma_{\mu}\gamma_5(-\hat k'_2+m_2)\right]=
-2\left[2k'_{\mu}(m_2-m_1)+P'_{\mu}(m_2+m_1)\right],
\\
\nonumber
&&T'_T=Tr\left[i\gamma_5
(\hat k'_1+m_1)i\sigma_{\mu\nu}(-\hat k'_2+m_2)\right]=
-4i\epsilon_{\mu\nu\alpha\beta}P'_{\alpha}
k'_{\beta},
\label{B.1}
\ee

and
\be
&&T_P=Tr\left[i\gamma_5(-\hat k_2+m_2)\gamma_5(\hat k_1+m_1)\right]=
2i(s-(m_2-m_1)^2)  ,
\\
\nonumber
&&T_A=Tr\left[i\gamma_5(-\hat k_2+m_2)i\gamma_{\mu}\gamma_5(\hat k_1+m_1)\right]=
2\left[2k_{\mu}(m_2-m_1)+P_{\mu}(m_2+m_1)\right]  ,
\\
\nonumber
&&T_T=Tr\left[i\gamma_5(-\hat k_2+m_2)i\sigma_{\mu\nu}(\hat k_1+m_1)\right]=
4i\epsilon_{\mu\nu\alpha\beta}P_{\alpha}k_{\beta} .
\label{B.2}
\ee

The convolutions of the traces $A_P=(T_P\, T'_P)$, $A_A=(T_A\, T'_A)$,
$A_T=(T_T\, T'_T)$ are equal to:
\be
A_P=&&-4(s-\Delta^2)(s'-\Delta^2) ,
\\
\nonumber
A_A=&&-16\Delta^2\sqrt{k^2_{\perp}k'^2_{\perp}}\,z
-4\frac{\sigma^2}{\sqrt{ss'}}
(s-\Delta^2)(s'-\Delta^2)   ,
\\
\nonumber
A_T=&&-32\sqrt{ss'}\sqrt{k^2_{\perp}k'^2_{\perp}}\,z .
\label{B.3}
\ee

\subsection{Traces for the $S=1$ states}

For the ($S=1$)-states, the traces are equal to:
\be
&&T'_S=Tr\left[\gamma_{\alpha'}^{\perp} (\hat k'_1+m_1)
(-\hat k'_2+m_2)\right]=4k'^{\perp}_{\alpha'}(m_2+m_1) ,
\\
\nonumber
&&T'_V=Tr\left[\gamma_{\alpha'}^{\perp}
(\hat k'_1+m_1)\gamma_{\mu}(-\hat k'_2+m_2)\right]
=2\left(g_{\alpha'\mu}^{\perp}(s'-(m_2-m_1)^2)+4k'^{\perp}_{\alpha'}
k'_{\mu}\right)  ,
\\
\nonumber
&&T'_A=Tr\left[\gamma_{\alpha'}^{\perp}
(\hat k'_1+m_1)i\gamma^{\perp}_{\mu}\gamma_5(-\hat k'_2+m_2)\right]=
4\epsilon_{\alpha'\mu\alpha\beta}k'_{\alpha}P'_{\beta}   ,
\\
\nonumber
&&T'_T=Tr\left[\gamma_{\alpha'}^{\perp}
(\hat k'_1+m_1)i\sigma_{\mu\nu}(-\hat k'_2+m_2)\right]=
\\
\nonumber
&&=2i\left[2(m_2-m_1)\left(g_{\alpha'\nu}^{\perp}k'_{\mu}
-g_{\alpha'\mu}^{\perp}k'_{\nu}\right)
+(m_1+m_2)\left(g_{\alpha'\nu}^{\perp}P'_{\mu}
-g_{\alpha'\mu}^{\perp}P'_{\nu}\right)\right],
\label{B.5}
\ee
and
\be
&&T_S=Tr \left[\gamma_{\beta'}^{\perp}(-\hat k_2+m_2)
\hat (k_1+m_1)\right] = 4k^{\perp}_{\beta'}(m_1+m_2) ,
\\
\nonumber
&&T_V=Tr\left[\gamma^{\perp}_{\beta'}(-\hat k_2+m_2)\gamma_{\mu}(\hat k_1+m_1)\right]=
2\left[g_{\mu\beta'}^{\perp}(s-(m_2-m_1)^2)+4k^{\perp}_{\beta'}k_{\mu}\right] ,
\\
\nonumber
&&T_A=Tr\left[\gamma_{\beta'}^{\perp}(-\hat k_2+m_2)i\gamma_{\mu}\gamma_5
(\hat k_1+m_1)\right]=
-4\epsilon_{\beta'\mu\alpha\beta}k_{\alpha}P_{\beta} ,
\\
\nonumber
&&T_T=Tr\left[\gamma_{\beta'}^{\perp}(-\hat k_2+m_2)i\sigma_{\mu\nu}
(\hat k_1+m_1)\right]=
\\
\nonumber
&&=2i\left[2(m_2-m_1)(g_{\beta'\mu}^{\perp}k_{\nu}
-g_{\beta'\nu}^{\perp}k_{\mu})
+(m_2+m_1)\left(g_{\beta'\mu}^{\perp}P_{\nu}
-g_{\beta'\nu}^{\perp}P_{\mu}\right)\right].
\label{B.6}
\ee

Corresponding convolution $B_S=(T_c\, T'_c)$ reads :
\be
&&(B_S)_{\beta'\alpha'}=16\,k^{\perp}_{\beta'}k'^{\perp}_{\alpha'}\sigma^2,
\\
\nonumber
&&(B_V)_{\beta'\alpha'}
=4\left[(g_{\beta'\alpha'}^{\perp}(s-\Delta^2)(s'-\Delta^2)
+16\,k^{\perp}_{\beta'}k'^{\perp}_{\alpha'}\sqrt{k^2_{\perp}k'^2_{\perp}}\,z +
\right .
\\
\nonumber
&&\left .
+4\, k^{\perp}_{\beta'}k^{\perp}_{\alpha'}(s'-\Delta^2)
+4\, k'^{\perp}_{\beta'}k'^{\perp}_{\alpha'}(s-\Delta^2)
+4\, k^{\perp}_{\beta'}k'^{\perp}_{\alpha'}
\frac{\sigma^2\Delta^2}{\sqrt{ss'}}
\right]       ,
\\
\nonumber
&&(B_A)_{\beta'\alpha'}=-16\sqrt{ss'}\left[k'^{\perp}_{\beta'} k^{\perp}_{\alpha'}-
g^{\perp}_{\beta'\alpha'}\sqrt{k^2_{\perp}k'^2_{\perp}}\, z\right] ,
\\
\nonumber
&& (B_T)_{\beta'\alpha'}=-8\left[4\Delta^2\left(k'^{\perp}_{\beta'}k^{\perp}_{\alpha'}
-g_{\beta'\alpha'}^{\perp}\sqrt{k^2_{\perp}k'^2_{\perp}}\, z\right)
-g_{\beta'\alpha'}^{\perp}
\frac{\sigma^2}{\sqrt{ss'}}(s-\Delta^2)(s'-\Delta^2)
\right]    .
\label{B.7}
\ee

\section{Appendix B: Convolutions of  trace factors}

Here we present the convolutions of the angular-momentum factors.
Let $z$ be:
\be
z=\frac{(k_{\perp}k'_{\perp})}{\sqrt{k_{\perp}^2}\sqrt{k_{\perp}'^2}}.
\ee

The convolutions for the ($S=1$) states read:
\be
X^{(J)}_{\mu_1 \mu_2 \cdots\mu_J}(k_{\perp})
X^{(J)}_{\mu_1 \mu_2 \cdots\mu_J}(k_{\perp}')=
\alpha (J)\left(\sqrt{k_{\perp}^2}\sqrt{k_{\perp}'^2}\right)^{J} P_{J}
(z). \label{C.1} \ee

Analogous convolutions for ($S=1$)-states are written as follows:
\be
&&X^{(J+1)}_{\mu_1 \mu_2 \cdots\mu_J\beta}(k_{\perp})
X^{(J+1)}_{\mu_1 \mu_2 \cdots\mu_J\alpha}(k_{\perp}')=
\frac{\alpha (J)}{J+1}\left(\sqrt{k_{\perp}^2}\sqrt{k_{\perp}'^2}\right)^{J}
\times
\\
\nonumber
&&\times\left[
 \frac{\sqrt{k_{\perp}'^2}}{\sqrt{k_{\perp}^2}}A_{P_{J,J+1}}(z)\, k_\beta^{\perp} k_\alpha^{\perp}
+\frac{\sqrt{k_{\perp}^2}}{\sqrt{k_{\perp}'^2}}B_{P_{J,J+1}}(z)\, k'^{\perp}_{\beta} k'^{\perp}_{\alpha}+
\right .
\\
\nonumber
&&\left .
+C_{P_{J,J+1}}(z)\, k_\beta^{\perp} k'^{\perp}_\alpha +D_{P_{J,J+1}}(z)\, k'^{\perp}_\beta
k_\alpha^{\perp} +\left(\sqrt{k_{\perp}^2}\sqrt{k_{\perp}'^2}\right)E_{P_{J,J+1}}(z)\,
g^{\perp}_{\beta\alpha} \right]\, ,
\label{C.2}
\ee

\be
&&X^{(J+1)}_{\mu_1 \mu_2 \cdots\mu_J\beta}(k_{\perp})
Z^{(J-1)}_{\mu_1 \mu_2 \cdots\mu_J,\alpha}(k_{\perp}')
= -\frac{\alpha (J)}{J}\frac{1}{k_{\perp}'^2}
\left(\sqrt{k_{\perp}^2}\sqrt{k_{\perp}'^2}\right)^{J}
\times
\\
\nonumber
&&\times\left[
 \frac{\sqrt{k_{\perp}'^2}}{\sqrt{k_{\perp}^2}}A_{P_{J,J+1}}(z)\, k_\beta^{\perp} k_\alpha^{\perp}
+\frac{\sqrt{k_{\perp}^2}}{\sqrt{k_{\perp}'^2}}
  \left(B_{P_{J,J+1}}(z)-(2J+1)A_J (z)\right) k'^{\perp}_\beta
  k'^{\perp}_\alpha+
\right .
\\
\nonumber
&&
+(C_{P_{J,J+1}}(z)-(2J+1)B_J (z))\, k_\beta^{\perp}
k'^{\perp}_\alpha+
\\
\nonumber
&&\left .
+D_{P_{J,J+1}}(z)\, k'^{\perp}_\beta k_\alpha^{\perp}
+\left(\sqrt{k_{\perp}^2}\sqrt{k_{\perp}'^2}\right)E_{P_{J,J+1}}(z)\,
g^{\perp}_{\beta\alpha} \right]\, ,
\label{C.3}
\ee
\be
&&Z^{(J-1)}_{\mu_1 \mu_2 \cdots\mu_J,\beta}(k_{\perp})
Z^{(J-1)}_{\mu_1 \mu_2 \cdots\mu_J,\alpha}(k_{\perp}')=
\frac{J+1}{J^2}\alpha (J)\left(\sqrt{k_{\perp}^2}\sqrt{k_{\perp}'^2}\right)^{J-2}
\times
\\
\nonumber
&&\times\left[
 \frac{\sqrt{k_{\perp}'^2}}{\sqrt{k_{\perp}^2}}
  \left(A_{P_{J,J+1}}(z)-(2J+1)A_J (z)\right)\, k_\beta^{\perp}
  k_\alpha^{\perp}+
\right .
\\
\nonumber
&&+\frac{\sqrt{k_{\perp}^2}}{\sqrt{k_{\perp}'^2}}
  \left(B_{P_{J,J+1}}(z)-(2J+1)A_J (z)\right)\, k'^{\perp}_\beta k'^{\perp}_\alpha
  +
\\
\nonumber
&&+\left(C_{P_{J,J+1}}(z)+\frac{(2J+1)^2}{J+1}P_J (z)-2(2J+1)B_J (z)
   \right)\, k_\beta^{\perp} k'^{\perp}_\alpha+
\\
\nonumber &&\left . +D_{P_{J,J+1}}(z)\, k'^{\perp}_\beta k_\alpha^{\perp}
+(\sqrt{k_{\perp}^2}\sqrt{k_{\perp}'^2})E_{P_{J,J+1}}(z)\, g^{\perp}_{\beta\alpha}
\right]\, ,
\label{C.4}
\ee

\be
&&\epsilon_{\beta \nu_1\nu_2\nu_3} P_{\nu_1}
    Z^{(J)}_{\nu_2 \mu_1\cdots\mu_J,\nu_3}(k_{\perp})\,\,
  \epsilon_{\alpha \lam_1\lam_2\lam_3} P'_{\lam_1}
    Z^{(J)}_{\lam_2 \mu_1\cdots\mu_J,\lam_3}(k_{\perp}')
\\
\nonumber
&&=\frac{(2J+3)^2}{(J+1)^3}\alpha(J)
\left(\sqrt{k_{\perp}^2}\sqrt{k_{\perp}'^2}\right)^{J-1}(PP')
\times
\\
\nonumber
&&\times
\left[-\sqrt{k_{\perp}^2}\sqrt{k_{\perp}'^2}\left((z^2-1)
D_{P_{J,J+1}}(z)+zE_{P_{J,J+1}}(z)\right)
g^{\perp}_{\beta\alpha}-
\right .
\\
\nonumber
&&-D_{P_{J ,J+1}}(z)
\left(\frac{\sqrt{k_{\perp}'^2}}{\sqrt{k_{\perp}^2}}k_{\beta}^{\perp}k_{\alpha}^{\perp}
+\frac{\sqrt{k_{\perp}^2}}{\sqrt{k_{\perp}'^2}}k'^{\perp}_{\beta} k'^{\perp}_{\alpha}
-z\,k_{\beta}^{\perp} k'^{\perp}_{\alpha}\right)+
\\
\nonumber
&&\left .
+\left(zD_{P_{J ,J+1}}(z)
+E_{P_{J,J+1}}(z)\right)k'^{\perp}_{\beta} k_{\alpha}^{\perp}\right]\, ,
\label{C.5}
\ee
and
\be
&&X^{(J)}_{\mu_1\cdots\mu_J}(k_{\perp})\,
\epsilon_{\alpha \nu_1\nu_2\nu_3} P'_{\nu_1}
Z^{(J)}_{\nu_2\mu_1\cdots\mu_J,\nu_3}(k_{\perp}')=
\\
\nonumber
&&=\frac{2J+3}{J+1}\alpha(J)
\left(\sqrt{k_{\perp}^2}\sqrt{k_{\perp}'^2}\right)^{J-1}A_{P_{J,J+1}}(z)\,\,
\epsilon_{\alpha P'kk'}\, .
\label{C.6}
\ee

Here
\be
&&A_{P_{J,J+1}}(z)=B_{P_{J,J+1}}(z)
=-\frac{2zP_J (z)+\left[Jz^2-(J+2)\right]P_{J+1}(z)}{(1-z^2)^2}\, ,
\\
\nonumber
&&C_{P_{J,J+1}}(z) =\frac{\left[(1-J)z^2+(J+1)\right]P_J (z)
+\left[(2J+1)z^2-(2J+3)\right]z P_{J+1} (z)}{(1-z^2)^2}\, ,
\\
\nonumber
&&D_{P_{J,J+1}}(z) =\frac{\left[(J+2)z^2-J\right]P_J
(z)-2zP_{J+1} (z)}{(1-z^2)^2}\, ,
\\
\nonumber
&&E_{P_{J,J+1}}(z)
=\frac{zP_J (z)-P_{J+1}(z)}{(1-z^2)}\, .
\label{C.7}
\ee

For the factors \\
$K_\beta X^{(J+1)}_{\mu_1 \mu_2 \cdots\mu_J\beta}(k_{\perp})
X^{(J+1)}_{\mu_1 \mu_2 \cdots\mu_J\alpha}(k_{\perp}') K_\alpha$, where $K=k,k'$,
 we need more complicated convolutions, namely:
:
\be
\nonumber
&&k_\beta\, X^{(J+1)}_{\mu_1 \mu_2 \cdots\mu_J\beta}(k_{\perp})
X^{(J+1)}_{\mu_1 \mu_2 \cdots\mu_J\alpha}(k_{\perp}')\, k_\alpha=
k_{\perp}^2 \alpha (J)\left(\sqrt{k_{\perp}^2}\sqrt{k_{\perp}'^2}\right)^{J+1}P_{J+1}(z)\, ,
\\
\nonumber
&&k_\beta\, X^{(J+1)}_{\mu_1 \mu_2 \cdots\mu_J\beta}(k_{\perp})
X^{(J+1)}_{\mu_1 \mu_2 \cdots\mu_J\alpha}(k_{\perp}')\, k'_\alpha=
\alpha (J)\left(\sqrt{k^2}\sqrt{k_{\perp}'^2}\right)^{J+2}P_{J}(z)\, ,
\\
\nonumber
&&k'_\beta\, X^{(J+1)}_{\mu_1 \mu_2 \cdots\mu_J\beta}(k_{\perp})
X^{(J+1)}_{\mu_1 \mu_2 \cdots\mu_J\alpha}(k_{\perp}')\, k'_\alpha=
k_{\perp}'^2 \alpha (J)\left(\sqrt{k_{\perp}^2}\sqrt{k_{\perp}'^2}\right)^{J+1}P_{J+1}(z)\, ,
\\
\nonumber
&&k'_\beta\, X^{(J+1)}_{\mu_1 \mu_2 \cdots\mu_J\beta}(k_{\perp})
X^{(J+1)}_{\mu_1 \mu_2 \cdots\mu_J\alpha}(k_{\perp}')\, k_\alpha=
\\
\nonumber
&&=\alpha (J)\left(\sqrt{k_{\perp}^2}\sqrt{k_{\perp}'^2}\right)^{J+2}
\left[\frac{2J+1}{J+1}zP_{J+1}(z)-\frac{J}{J+1}P_{J}(z)\right]\, ,
\\
\nonumber
&&g^{\perp}_{\beta\alpha}\,
X^{(J+1)}_{\mu_1 \mu_2 \cdots\mu_J\beta}(k_{\perp})
X^{(J+1)}_{\mu_1 \mu_2 \cdots\mu_J\alpha}(k_{\perp}')=
\frac{2J+1}{J+1}\alpha (J)
\left(\sqrt{k_{\perp}^2}\sqrt{k_{\perp}'^2}\right)^{J+1}P_{J+1}(z)\, ,
\label{C.10}
\ee

as well as  for the factors
$K_\beta X^{(J+1)}_{\mu_1 \mu_2 \cdots\mu_J\beta}(k_{\perp})
Z^{(J-1)}_{\mu_1 \mu_2 \cdots\mu_J,\alpha}(k_{\perp}') K_\alpha$:
\be
\nonumber
&&k_\beta\, X^{(J+1)}_{\mu_1 \mu_2 \cdots\mu_J\beta}(k_{\perp})
Z^{(J-1)}_{\mu_1 \mu_2 \cdots\mu_J,\alpha}(k_{\perp}')\, k_\alpha=
k_{\perp}^4 \alpha (J)\left(\sqrt{k_{\perp}^2}\sqrt{k_{\perp}'^2}\right)^{J-1}P_{J-1}(z)\, ,
\\
\nonumber
&&k_\beta\, X^{(J+1)}_{\mu_1 \mu_2 \cdots\mu_J\beta}(k_{\perp})
Z^{(J-1)}_{\mu_1 \mu_2 \cdots\mu_J,\alpha}(k_{\perp}')\, k'_\alpha=
k_{\perp}^2\alpha (J)\left(\sqrt{k_{\perp}^2}\sqrt{k_{\perp}'^2}\right)^{J}P_{J}(z)\, ,
\\
\nonumber
&&k'_\beta\, X^{(J+1)}_{\mu_1 \mu_2 \cdots\mu_J\beta}(k_{\perp})
Z^{(J-1)}_{\mu_1 \mu_2 \cdots\mu_J,\alpha}(k_{\perp}')\, k'_\alpha=
\alpha (J)\left(\sqrt{k_{\perp}^2}\sqrt{k_{\perp}'^2}\right)^{J+1}P_{J+1}(z)\, ,
\\
\nonumber
&&k'_\beta\, X^{(J+1)}_{\mu_1 \mu_2 \cdots\mu_J\beta}(k_{\perp})
Z^{(J-1)}_{\mu_1 \mu_2 \cdots\mu_J,\alpha}(k_{\perp}')\, k_\alpha=
k_{\perp}^2\alpha (J)\left(\sqrt{k_{\perp}^2}\sqrt{k_{\perp}'^2}\right)^{J}P_{J}(z)\, ,
\\
\nonumber
&&g^{\perp}_{\beta\alpha}\, X^{(J+1)}_{\mu_1 \mu_2 \cdots\mu_J\beta}(k_{\perp})
Z^{(J-1)}_{\mu_1 \mu_2 \cdots\mu_J,\alpha}(k_{\perp}')=0\, ,
\label{C.11}
\ee

and  for the factors $K_\beta Z^{(J-1)}_{\mu_1 \mu_2 \cdots\mu_J,\beta}(k_{\perp})
Z^{(J-1)}_{\mu_1 \mu_2 \cdots\mu_J,\alpha}(k_{\perp}') K_\alpha$:

\be
\nonumber
&&k_\beta\, Z^{(J-1)}_{\mu_1 \mu_2 \cdots\mu_J,\beta}(k_{\perp})
Z^{(J-1)}_{\mu_1 \mu_2 \cdots\mu_J,\alpha}(k_{\perp}')\, k_\alpha=
k_{\perp}^2 \alpha (J)\left(\sqrt{k_{\perp}^2}\sqrt{k_{\perp}'^2}\right)^{J-1}P_{J-1}(z)\, ,
\\
\nonumber
&&k_\beta\, Z^{(J-1)}_{\mu_1 \mu_2 \cdots\mu_J,\beta}(k_{\perp})
Z^{(J-1)}_{\mu_1 \mu_2 \cdots\mu_J,\alpha}(k_{\perp}')\, k'_\alpha=
\alpha (J)\left(\sqrt{k_{\perp}^2}\sqrt{k_{\perp}'^2}\right)^{J}P_{J}(z)\, ,
\\
\nonumber
&&k'_\beta\, Z^{(J-1)}_{\mu_1 \mu_2 \cdots\mu_J,\beta}(k_{\perp})
Z^{(J-1)}_{\mu_1 \mu_2 \cdots\mu_J,\alpha}(k_{\perp}')\, k'_\alpha=
k_{\perp}'^2 \alpha (J)\left(\sqrt{k_{\perp}^2}\sqrt{k_{\perp}'^2}\right)^{J-1}P_{J-1}(z)\, ,
\\
\nonumber
&&k'_\beta\, Z^{(J-1)}_{\mu_1 \mu_2 \cdots\mu_J,\beta}(k_{\perp})
Z^{(J-1)}_{\mu_1 \mu_2 \cdots\mu_J,\alpha}(k_{\perp}')\, k_\alpha
\\
\nonumber
&&=\alpha (J)\left(\sqrt{k_{\perp}^2}\sqrt{k_{\perp}'^2}\right)^{J}
\left[\frac{2J+1}{J}zP_{J-1}(z)-\frac{J+1}{J}P_{J}(z)\right]\, ,
\\
\nonumber
&&g^{\perp}_{\beta\alpha}\,
Z^{(J-1)}_{\mu_1 \mu_2 \cdots\mu_J,\beta}(k_{\perp})
Z^{(J-1)}_{\mu_1 \mu_2 \cdots\mu_J,\alpha}(k_{\perp}')=
\frac{2J+1}{J}\alpha (J)
\left(\sqrt{k_{\perp}^2}\sqrt{k_{\perp}'^2}\right)^{J-1}P_{J-1}(z)\, ,
\label{C.12}
\ee

and for the factors $K_\beta \epsilon_{\beta \nu_1\nu_2\nu_3} P_{\nu_1}
Z^{(J)}_{\nu_2 \mu_1\cdots\mu_J,\nu_3}(k_{\perp})\,\,
\epsilon_{\alpha \lam_1\lam_2\lam_3} P'_{\lam_1}
Z^{(J)}_{\lam_2 \mu_1\cdots\mu_J,\lam_3}(k_{\perp}') K_\alpha$:
\be
\nonumber
&&k_\beta\, \epsilon_{\beta \nu_1\nu_2\nu_3} P_{\nu_1}
Z^{(J)}_{\nu_2 \mu_1\cdots\mu_J,\nu_3}(k_{\perp})\,\,
\epsilon_{\alpha \lam_1\lam_2\lam_3} P'_{\lam_1}
Z^{(J)}_{\lam_2 \mu_1\cdots\mu_J,\lam_3}(k_{\perp}')\, k_\alpha=0\, ,
\\
\nonumber
&&k_\beta\, \epsilon_{\beta \nu_1\nu_2\nu_3} P_{\nu_1}
Z^{(J)}_{\nu_2 \mu_1\cdots\mu_J,\nu_3}(k_{\perp})\,\,
\epsilon_{\alpha \lam_1\lam_2\lam_3} P'_{\lam_1}
Z^{(J)}_{\lam_2 \mu_1\cdots\mu_J,\lam_3}(k_{\perp}')\, k'_\alpha=0\, ,
\\
\nonumber
&&k'_\beta\, \epsilon_{\beta \nu_1\nu_2\nu_3} P_{\nu_1}
Z^{(J)}_{\nu_2 \mu_1\cdots\mu_J,\nu_3}(k_{\perp})\,\,
\epsilon_{\alpha \lam_1\lam_2\lam_3} P'_{\lam_1}
Z^{(J)}_{\lam_2 \mu_1\cdots\mu_J,\lam_3}(k_{\perp}')\, k'_\alpha=0\, ,
\\
\nonumber
&&k'_\beta\, \epsilon_{\beta \nu_1\nu_2\nu_3} P_{\nu_1}
Z^{(J)}_{\nu_2 \mu_1\cdots\mu_J,\nu_3}(k_{\perp})\,\,
\epsilon_{\alpha \lam_1\lam_2\lam_3} P'_{\lam_1}
Z^{(J)}_{\lam_2 \mu_1\cdots\mu_J,\lam_3}(k_{\perp}')\, k_\alpha
\\
\nonumber
&&=\frac{(2J+3)^2}{(J+1)^3}\alpha(J)
\left(\sqrt{k_{\perp}^2}\sqrt{k_{\perp}'^2}\right)^{J+1}(PP')
\left[zP_J(z)-P_{J+1}(z)\right]\, ,
\\
\nonumber
&&g^{\perp}_{\beta\alpha}\,\epsilon_{\beta \nu_1\nu_2\nu_3} P_{\nu_1}
Z^{(J)}_{\nu_2 \mu_1\cdots\mu_J,\nu_3}(k_{\perp})\,\,
\epsilon_{\alpha \lam_1\lam_2\lam_3} P'_{\lam_1}
Z^{(J)}_{\lam_2 \mu_1\cdots\mu_J,\lam_3}(k_{\perp}')=
\\
\nonumber
&&=-\frac{J(2J+3)^2}{(J+1)^3}\alpha(J)
\left(\sqrt{k_{\perp}^2}\sqrt{k_{\perp}'^2}\right)^{J}(PP')
P_J(z)\, .
\label{C.13}
\ee

\end{document}